\documentclass[runningheads]{llncs}
\usepackage{graphicx}
\usepackage{amsmath}
\usepackage{hyperref}
\usepackage{algorithm}
\usepackage{algpseudocode}
\usepackage{graphicx}
\usepackage{subcaption}
\usepackage[export]{adjustbox}

\begin{document}
\title{Being Automated or Not? Risk Identification of Occupations with Graph Neural Networks}

%
%
\author{Dawei Xu\inst{1} \and
Haoran Yang\inst{1} \and
Marian-Andrei Rizoiu\inst{2} \and
Guandong Xu\inst{2}}
%
%
\institute{University of Technology Sydney, Sydney, Australia \\
\email{\{dawei.xu, haoran.yang-2\}@student.uts.edu.au}\\
\and
University of Technology Sydney, Sydney, Australia\\
\email{\{marian-andrei.rizoiu, guandong.xu\}@uts.edu.au}}
%
\titlerunning{Occupation Automated Risk Identification}
\maketitle              
\begin{abstract}
The rapid advances in automation technologies, such as artificial intelligence (AI) and robotics, pose an increasing risk of automation for occupations, with a likely significant impact on the labour market.
Recent social-economic studies suggest that nearly 50\% of occupations are at high risk of being automated in the next decade. However, the lack of granular data and empirically informed models have limited the accuracy of these studies and made it challenging to predict which jobs will be automated. In this paper, we study the automation risk of occupations by performing a classification task between automated and non-automated occupations. The available information is 910 occupations' task statements, skills and interactions categorised by Standard Occupational Classification (SOC). To fully utilize this information, we propose a graph-based semi-supervised classification method named \textbf{A}utomated \textbf{O}ccupation \textbf{C}lassification based on \textbf{G}raph \textbf{C}onvolutional \textbf{N}etworks (\textbf{AOC-GCN}) to identify the automated risk for occupations. This model integrates a heterogeneous graph to capture occupations' local and global contexts. The results show that our proposed method outperforms the baseline models by considering the information of both internal features of occupations and their external interactions. This study could help policymakers identify potential automated occupations and support individuals' decision-making before entering the job market.

\end{abstract}
\section{Introduction}
In light of rapid development in the fields of artificial intelligence and robotics, the number of jobs for certain occupations is decreasing alarmingly. Also, massive applications of automation technologies, especially in the workplace, are raising fears of large-scale technological unemployment and a renewed appeal for policymakers to handle the consequences of technological change. From a historical perspective, every disappearance of occupations comes with increasing unemployment \cite{ref_1,Dawson2021} which brings individual existence crisis, social instability, and techno-phobia that may hinder the future development of science and technologies. Therefore, occupations' automated risk identification is critical to both policymakers and individuals.\par
Generally, an occupation at risk of being automated has attributes containing routine and rule-based information, which are prone to be replaced by automation technologies \cite{ref_4,ref_5}. Predicting the automated risk of occupations can be performed by capturing the routine and rule-based features in their attributes, more specifically, their tasks and skills information \cite{ref_3}. Therefore, this prediction task can be conducted as a binary classification between automated and non-automated occupations by mining their internal attributes and external interactions.\par
\vspace{-0.5cm}
\begin{figure}
  \includegraphics[width=\linewidth]{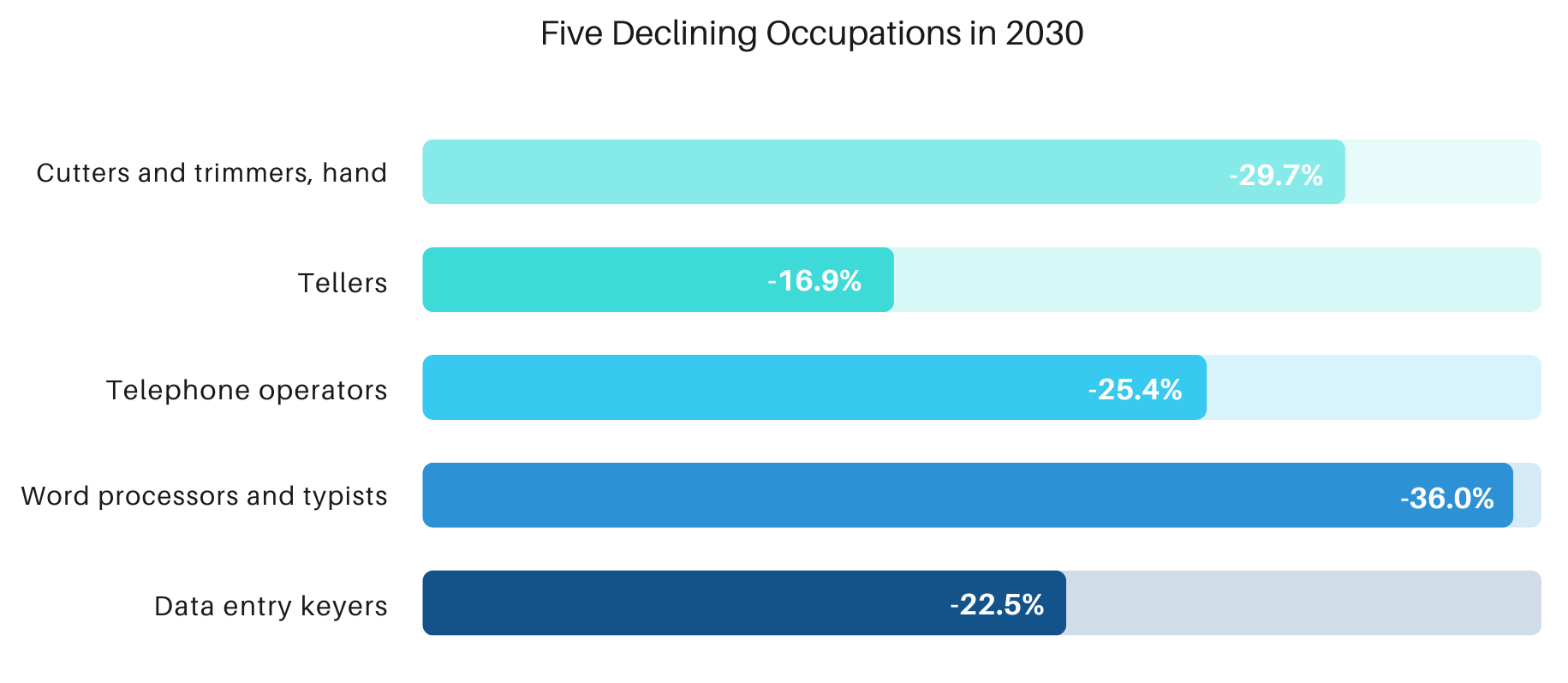}
  \caption{BLS Projected Declining Occupations in 2030}
  \label{fig:introduction}
\end{figure}
\vspace{-0.5cm}
However, this task becomes more challenging due to the following three characteristics of the data:
\begin{description}
\setlength{\itemindent}{0in}
  \item[$\bullet$ Limited Data Source.] Predicting automated risk of occupations requires tasks and skills data that are at their most granular level to reflect micro differences among occupations. Standard labor data from U.S. Bureau of Labor Statistics (BLS) or Australian Bureau of Statistics (ABS) focus on demographic, economic or other aggregate statistics that lack resolutions into the defining features to distinguish different occupations. Extracting granular tasks and skills data from job recruitment platforms is also not feasible since they are dynamic, computing expensive and not all jobs are posted online. The numbers of job posts are also various across occupations which may lead to imbalanced data \cite{ref_7}. Furthermore, the lack of ground truth data is also a barrier hindering the prediction of automated risk.
  \item[$\bullet$ Non-numerical Data Type.] Most of the tasks and skills data are in text form, which are single words, grouped words or sentences. These text data cannot be directly processed by statistical models. Current studies \cite{ref_1}\cite{ref_5} use importance or level value attached to each task or skill as its numerical representations; however, this kind of representation is aggregated and coarse which may obfuscate the differential impact of various technologies and skill requirements then further affects models' performance in distinguishing occupations. Therefore, choosing a method that can not only encode the text data into a numerical format but maintain the features hidden in the context are necessary.
  \item[$\bullet$ Complex Data Structure.] Since occupations are likely to interact by attributes, e.g., they may share similar skillsets, the prediction of automated risk cannot be limited to occupations' internal attributes but also the connections with other occupations \cite{ref_3}. Moreover, various attributes may lead to different data dimension sizes, and the relationships among occupations can be multiple paths \cite{Dawson2019}. Therefore, linear data structures are not capable of capturing all the information, and a more complicated non-linear structure and computational model should be designed to recognise the patterns in the information.
\end{description}

Moreover, existing methods designed for the task have several limitations. For instance, current studies conduct experiments on BLS wages and education data \cite{ref_27} or the importance level, a numerical value of task statements \cite{ref_1}\cite{ref_2}, and they utilise linear data structure to represent the various attributes. According to the defects mentioned above, the accuracy of their results is unavoidable compromised. A better solution is to use Natural Language Processing (NLP) technologies to generate numerical embeddings for tasks and skills text data and use a non-linear data structure such as a graph to capture not only the features of occupations but also the interactions between occupations.
Therefore, we propose a semi-supervised classification method named Automated Occupation Classification based on Graph Convolutional Networks (AOC-GCN) to identify the automated risk for occupations. We will be using tasks and skills datasets extracted from the O*NET database, which is a publicly available occupational information network containing hundreds of job definitions, to understand the labour market.\par

In summary, the contributions of our work are listed below:
\begin{description}
\setlength{\itemindent}{0in}
    \item[$\bullet$] We use Natural Language Processing (NLP) technologies -- Word2Vec and Doc2Vec to generate embeddings for tasks and skills data, which is, to the best of our knowledge, the first work in this domain using NLP to generate numerical representations for text data. 
    \item[$\bullet$] We propose a GCN-based heterogeneous graph classification method that aggregates both node features and network-wide behaviours, which is also the first work in this domain.
    \item[$\bullet$] Our proposed method significantly outperforms baseline models, and the results are verified through comparison with government statistical data.
\end{description}\par
The rest of the paper is organized as follows. Section \ref{section:relate} lists the related work. We elaborate on the proposed AOC-GCN method in Section \ref{section:meth}. Section \ref{section:experiment} compares and analyzes the experimental results of our method. In Section \ref{section:result}, we discuss the results and compare them with government statistical data. In Section \ref{section:limit}, we discuss limitations and future work, and the paper is concluded in Section \ref{section:conclusion}.\par
\section{Related Work}\label{section:relate}

\subsection {Occupations Automated Risk identification.}
Most existing studies about automated occupation identification are based on expert-based assessments, or basic statistical models focusing on probability space. Frey et al. \cite{ref_1} assessed the automated probability of occupations using a Gaussian process classification and concluded that 47\% of current US employment is at high risk of being automated. Similar studies have been conducted at the impact of automation on occupations in other countries and reached cautionary conclusions: automated occupations rate will be 35\% in Finland \cite{ref_9}, 59\% in Germany \cite{ref_8}, and 45 to 60 \% across Europe \cite{ref_10}. On the contrary, Arntz et al. \cite{ref_2} followed the experiments in \cite{ref_1} but used job-level data and a different coefficient estimates method, and concluded that a less alarming 9\% of employment is at risk. Vermeulen et al. \cite{ref_5} and Arntz et al. \cite{ref_11} also conducted job-level automatability scorings and get similar findings that about 5\% of jobs have automated risks. However, the data utilized in Arntz et al. are job survey data conducted on different workers. Arntz et al. considers this data as job level data compared to occupation level data without considering the duplicates of same tasks and skills performed by different workers' job. Therefore, the results may be underestimated. That is to say, job-level studies consider each task to be independent and unique which masked the similarity between tasks. Furthermore, current studies failed to discover the semantic information hidden in text data and ignored the interactions between occupations.\par

\subsection {Graph Neural Networks.}
Recent years have witnessed a growing application in using graph neural network-based algorithms to solve graph structure problems~\cite{Mihaita2019,Mihaita2019a,Mihaita2020}. These methods include both supervised \cite{ref_6} and unsupervised methods \cite{ref_14} \cite{ref_15}. Especially, Graph Convolutional Networks (GCN) \cite{ref_6} have achieved significant performance compared to previous methods, which is an efficient variant of Convolutional Neural Networks (CNN) \cite{ref_17} operating directly on graphs. In GCN, both the graph structure information and node features are also aggregated from neighbours during convolution. The graph convolution operation is defined as feature aggregations of neighbours, and through iterative convolutions, the whole graph's information can be propagated and aggregated to each part of the graph. After GCN, GraphSAGE \cite{ref_16} is proposed, which is a simpler but efficient inductive learning model that breaks the limitation of applying GCN in transductive learning. Furthermore, Graph Attention Networks (GAT) \cite{ref_18} introduces an attention mechanism into GCN and allows nodes to focus on the most relevant neighbours during training.\par
In this paper, a GCN-based method is first applied to automated occupation classification task through building an occupation-skill graph. Our proposed method can encode not only the features of occupation but also the interaction between occupations.\par
\section{Methodology}\label{section:meth}
\subsection{Preliminaries}
Generally, GCN-based models \cite{ref_6}\cite{ref_16}\cite{ref_18} consist of multiple propagation layers and all the nodes are updated simultaneously in each propagation layer. A propagation layer can be divided into two sub-layers: $aggregation$ and $combination$. Let $ G =  (V,\,E) $ be a graph with node $ v \in V $, edge $ (v,\,v^\prime) \in E $, and node feature $ x_v = h_v^0 \in \textbf{R}^{d_0} $ for $ v \in V $ where ${d_0}$ denotes the feature dimension of the node and ${h_v^l} \in {\textbf{R}}^{d_l}$ denotes the hidden state of node $v$ learned by the $l$-th layer of the model. For a GCN with $L$ layers, aggregation and combination sub-layers at $l$-th layer($l$ = 1, 2, ... $L$) can be written as :
\begin{equation}
{h_{N(v)}^l} = \sigma(W^l \cdot \operatorname{AGG}(\{{h_{v^\prime}^{l-1}, \forall v^\prime} \in {N(v)}\}))
\end{equation}
\begin{equation}
{h_v^l} = \operatorname{COMBINE}(h_v^{l-1}, h_{N(v)}^l)
\end{equation}
where $N(v)$ is a set of nodes adjacent to $v$, $AGG$ is a function used to aggregate embeddings from neighbor nodes of $v$. ${W^l}$ is a trainable matrix shared among all nodes at layer $l$. $\sigma$ is a non-linear activation function e.g., $RELU$. $h_{N(v)}^l$ denotes the aggregated feature of all neighbors of node $v$ at $l$-th layer. $COMBINE$ function is used to combine the embedding of node itself and the aggregated embeddings of neighbors.\par
For heterogeneous graph with different types of nodes and edges, it always comes with different sizes of embeddings. For each edge connecting two different nodes, we generally concatenate the edge embedding from the last propagation layer together with embeddings of the two nodes this edge links to. Similarly, for each node, besides the information from neighbor nodes, the features of edges connected to them are also collected. Therefore the aggregation sub-layer for an edge $e$ can be defined as:
\begin{equation}
{h_e^l} = \sigma(W_E^l \cdot \operatorname{AGG}_E^l(h_e^{l-1}, h_{U(e)}^{l-1}, h_{I(e)}^{l-1}))
\end{equation}
where $h_{U(e)}^{l-1}$ and $h_{I(e)}^{l-1}$ denote the hidden states of two types of nodes $U$ and $I$ from the previous propagation layer and
\begin{equation}
\operatorname{AGG}_E^l(h_e^{l-1}, h_{U(e)}^{l-1}, h_{I(e)}^{l-1}) = \operatorname{concat}(h_e^{l-1}, h_{U(e)}^{l-1}, h_{I(e)}^{l-1})
\end{equation}\par
For the two nodes $u \in U$ and $i \in I$, the aggregated neighbor embedding $h_{N(u)}^l$ and $h_{N(i)}^{l}$ can be calculated as:
\begin{equation}
\begin{split}
& {h_{N(u)}^l} = \sigma(W_U^l \cdot \operatorname{AGG}_U^l(\operatorname{concat}(h_i^{l-1}, h_e^{l-1}), \forall e = (u, i) \in E(u))) \\
& {h_{N(i)}^l} = \sigma(W_I^l \cdot \operatorname{AGG}_I^l(\operatorname{concat}(h_u^{l-1}, h_e^{l-1}), \forall e = (u, i) \in E(i)))
\end{split}
\end{equation}\par
Here the two types of nodes maintain different parameters $(W_U^l, W_I^l)$ and different aggregation functions $(AGG_U^l, AGG_I^l)$.\par
After aggregating the neighbors' information, we follow the concatenation in [6] for the two types of nodes as
\begin{equation}
\begin{split}
& {h_u^l} = \operatorname{concat}(V_U^l \cdot h_u^{l-1}, h_{N(u)}^l) \\
& {h_i^l} = \operatorname{concat}(V_I^l \cdot h_i^{l-1}, h_{N(i)}^l)
\end{split}
\end{equation}
Where $V_U^l$ and $V_U^l$ denote trainable weight matrix for node $U$ and node $I$, and the $h_u^l$ and $h_i^l$ are the hidden states of $l$-th layer for the two types of nodes.
\subsection{Problem Setup}
Our purpose is to identify the automated risk for occupations listed in Standard Classification Code (SOC). The attributes used in this problem are task statements and skills extracted from O*NET database. According to the properties of data, each task statement is unique for each occupation while each skill data are a list of skills which are overlapped and shared among other occupations. So we can build a bipartite graph $G(O,\,S,\,E)$ where $O$ is the set of occupation nodes, $S$ is the set of skill nodes, and $E$ is the set of edges between occupations and skills. An edge $e \in E$ from an occupation $o \in O$ to a skill $s \in S$ exists if $o$ contains $s$. It is worth noting that the task statements data will be used as occupation nodes' feature representations which will be introduced in section \ref{section:feature}. Therefore, the task can be formulated to a node classification problem on an undirected bipartite graph with two types of attributed nodes.
\subsection{Graph Convolutional Networks on Occupation-skill Graph}\label{section:node}
The occupation-skill graph is a bipartite graph with two types of nodes and one type of edge. Moreover, there is no attributes on edges in our problem, so applying a heterogeneous GCN to solve this problem can be excessive. In fact, we take a shortcut way by applying GCN to treat this bipartite graph as a homogeneous graph by setting identical dimension sizes for both nodes features. Therefore, the propagation rule at layer $l$ is this problem can be defined as:
\begin{equation}
{h_v^l} = \operatorname{COMBINE}(h_v^{l-1}, \sigma(W^l \cdot \operatorname{AGG}(\{{h_{v^\prime}^{l-1}, \forall v^\prime} \in {N(v)}\})))
\end{equation}
where $v \in O\cup S$, $N(v)$ is a set of nodes adjacent to $v$, $h_v^l$ denote the hidden states at $l$-th layer for each nodes.
\subsection{Nodes Feature Representation}\label{section:feature}
Text data of task statements and skills should be converted into an embedding before being encoded as nodes' features. Here we apply two word embeddings technologies -- Word2vec \cite{ref_19} and Doc2vec \cite{ref_20} to generate feature representations for occupation nodes and skill nodes. Firstly, we use Glove pretrained word embeddings to initialize the Word2vec model and retrain the model on task statements data to aggregate the context and semantic similarity information in this domain. The output from the Word2vec model is a word embeddings table containing vectors for each word. Since the skill data are single word or grouped words, we represent each skill's feature by simply applying the word vector or making additions of multiple word vectors from word embeddings. Then, we initialize our Doc2vec model with word embeddings from last step and train on task statements data to generate document-level embeddings as occupation nodes feature representations. In detail,
\begin{equation}
{h_s^0} = \operatorname{fusion}(\textbf{\textit{w}}_1, \textbf{\textit{w}}_2, ..., \textbf{\textit{w}}_n)
\end{equation}
\begin{equation}
{h_o^0} = \operatorname{doc2vec}(t_0)
\end{equation}
where ${h_s^0}$, ${h_o^0}$ denote the initial embedding of a skill node and an occupation node. $\textbf{\textit{w}}_i$ represents the $i$-th word vector in a skill generated from Word2vec. $t_0$ is the task statement document of an occupation. Therefore, the parameters of Word2vec and Doc2vec are trained together with others in the model described in Section \ref{section:node}.
\subsection{Semi-supervised Automated Occupation Classification Model}
The final embeddings of AOC-GCN are the concatenation of nodes embeddings learned from Occupation-skill Graph, so we can get
\begin{equation}
y = \operatorname{softmax}(f(\operatorname{concat}(z_o, z_s)))
\end{equation}
where $z_o$ and $z_s$ denote the embeddings of O and S learned by the proposed GCN method. $f(\cdot)$ is the mapping function to map the embeddings to lower dimension space before putting into $softmax$ which is the classifier used to classify automated and non-automated occupations. This method is semi-supervised because the whole graph is updated during training process which is on a partially labeled dataset. The whole pipeline of the study is shown as in Fig \ref{fig:flow}.
\begin{figure}
\centering
    \begin{subfigure}[b]{\textwidth}            
            \includegraphics[width=\textwidth]{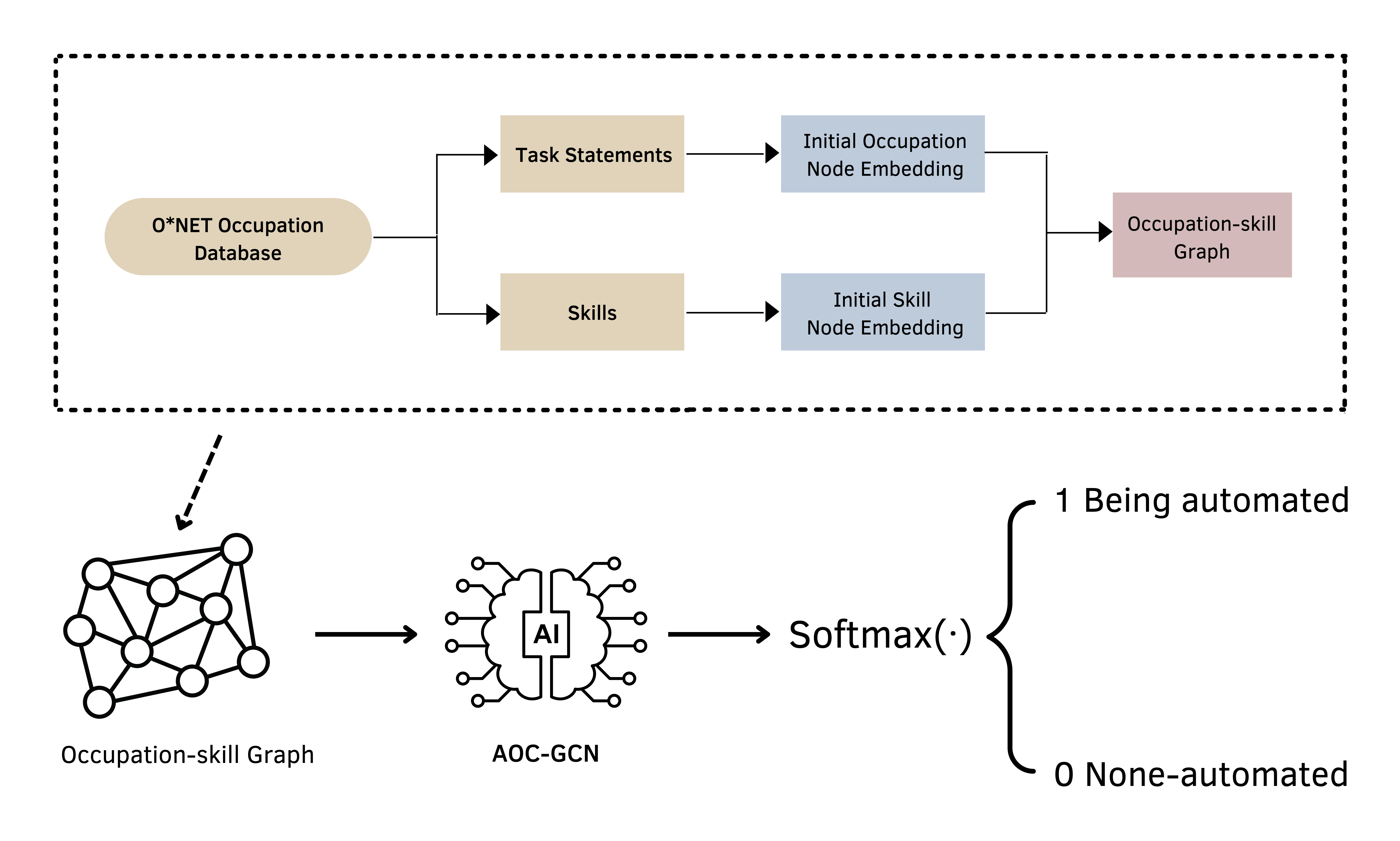}
            \label{fig:flow}
    \end{subfigure}%
    \caption{The whole System Pipeline}\label{fig:flow}
\end{figure}


\section{Experiment}\label{section:experiment}
\subsection{Dataset and Metrics}
Experiments are conducted on occupation lists, task statements and skills in Standard Classification Code (SOC) from O*NET database. The label information utilized in the experiments is generated from expert-based assessments on O*NET occupations conducted by \cite{ref_1}\cite{ref_5} with automated occupations being labeled as 1 and non-automated occupations being labeled as 0. However, the original label information has defects that the number of two classes are unbalanced, and some occupations are too close e.g., web developer and software programmer. We preprocess the data and finally select 112 labeled occupations with 56 automated occupations and 56 non-automated occupations which also have distance between each other. Then, all the data will form an undirected graph structure with 910 occupation nodes, 135 skill nodes and 13222 edges. Training, validation and test set are randomly split with a ratio of 8:1:1. Given the prediction and ground truth, we evaluate the model's performance using accuracy, precision, recall and F1 score.
\subsection{Baselines}
We compare our model with traditional classification models. Specifically, we design features of each occupation by concatenate the embeddings generated from Section \ref{section:feature} for both task statements and skills. Then we use zero padding to make sure all the occupations have identical feature dimension sizes before putting into baseline models. The chosen baseline models are as following:\par
$\bullet$ \textbf{Decision Tree}. DT splits the dataset as a tree based on a set of rules and conditions which is a supervised learning algorithm that can be used for both classification and regression \cite{ref_24}.\par
$\bullet$ \textbf{Random Forest}. Random Forest is an ensemble method which combines the output of multiple decision trees to reach a single result \cite{ref_22}. \par
$\bullet$ \textbf{Adaptive Boost}. AdaBoost (Adaptive Boosting) is a boosting technique that aims at combining multiple weak classifiers to build one strong classifier and is widely used in both classification and regression problem \cite{ref_21}.\par
$\bullet$ \textbf{Light Gradient Boosting Machine}. LightGBM is a gradient boosting framework based on decision trees to increase the efficiency of the model and reduce memory usage \cite{ref_23}.\par


\subsection{Results and Analysis}
\subsubsection {Evaluation.} We evaluate the results of AOC-GCN model on O*NET dataset with above metrics and compare them with baseline models. The results of the comparison are shown in Table \ref{tab1} below.\par
\begin{table}
\centering
\caption{Preprocessed results of O*NET dataset}\label{tab1}
\begin{tabular}{|l l l l l|}
\hline
{\bfseries Method\hspace{2cm}}  & {\bfseries Accuracy\hspace{0.5cm}} & {\bfseries Precision\hspace{0.5cm}} & {\bfseries Recall\hspace{1cm}} & {\bfseries F1\hspace{0.5cm}}\\
\hline
{Decision Tree}  & {0.7143} & {0.6667} & {0.8571} & {0.7500}\\

{Random Forest}  & {0.7500} & {0.7692} & {0.7143} & {0.7407}\\

{LightGBM}  & {0.7500} & {0.6842} & {0.9286} & {0.7879}\\

{AdaBoost}  & {0.8571} & {0.8125} & {0.9286} & {0.8667}\\

{AOC-GCN}  & {\textbf{0.9091}} & {\textbf{1.0000}} & {\textbf{0.8750}} & {\textbf{0.9333}}\\
\hline
\end{tabular}
\end{table}
We find that our AOC-GCN method performs significantly better than other baseline models, since our model not only uses external node features, but also capture the graph structural information which is the interactions between nodes. Besides, baseline models using raw features only already achieve good results, the GCN's combination of both semantic and structural information of relations successfully attains the best performance. Furthermore, our proposed model ensures high model precision to avoid disturbing non-automated occupations.

\subsubsection {Parameter Experiments.} In this section, we explore the effects on the model performance of different parameters. The key parameters used here are the dimensions of initial node embedding sizes and the dimension sizes of GCN layer. The dimension sizes of node embeddings range from 50 to 300 and the dimension sizes of GCN layer range from 16 to 512.
\begin{figure}
    \begin{subfigure}[b]{0.48\textwidth}  
            \includegraphics[width=\textwidth]{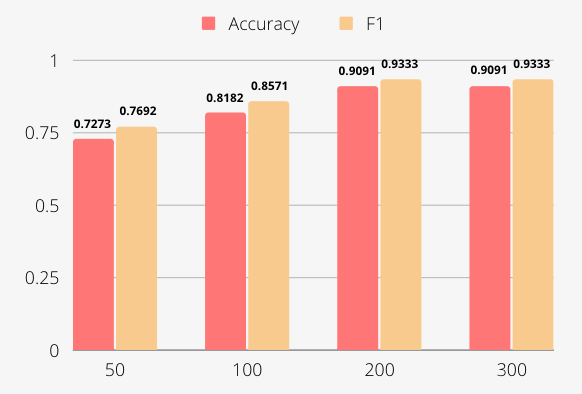}
            \caption{Model performance on different \\initial node embeddings size}
            \label{fig:SRl}
    \end{subfigure}%
    \begin{subfigure}[b]{0.48\textwidth}
            \hspace{1em}
            \includegraphics[width=\textwidth]{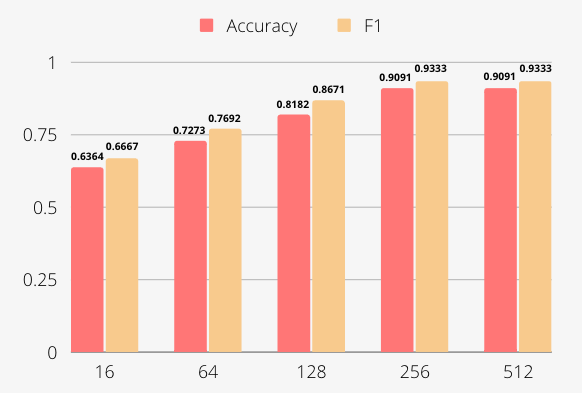}
            \caption{Model performance on different \\GCN layer size}
            \label{fig:D-Imager}
    \end{subfigure}
    \caption{Parameters Effect On Model Performance}\label{fig:para}
\end{figure}

We can see from the Fig \ref{fig:para} that with the increase of sizes of initial node embeddings from 50 to 200, both the accuracy and F1 scores rise and reach the maximum value. When the size is greater than 200, both of the values keep stable. The same pattern can be found in GCN layer dimension sizes, when the size increases from 16 to 256, both of the accuracy and F1 increase. When the size is greater than 256, the two scores keep unchanged. Therefore, we can conclude that the model gets the best performance when the initial node embeddings size is greater than 200 and the GCN layer dimension size is greater than 256.\par
\subsubsection{Graph Embeddings Visualization}
\begin{figure}
\centering
    \begin{subfigure}[b]{0.6\textwidth}            
            \includegraphics[width=\textwidth]{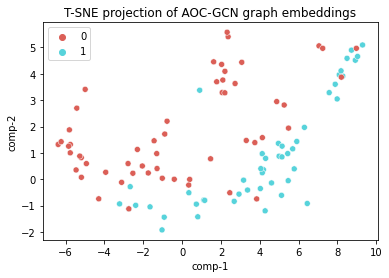}
            \label{fig:SRl}
    \end{subfigure}%
    \caption{t-SNE projection of GCN embeddings}\label{fig:tsne_16}
\end{figure}

\begin{figure}
\centering
    \begin{subfigure}[b]{0.7\textwidth}            
            \includegraphics[width=\textwidth]{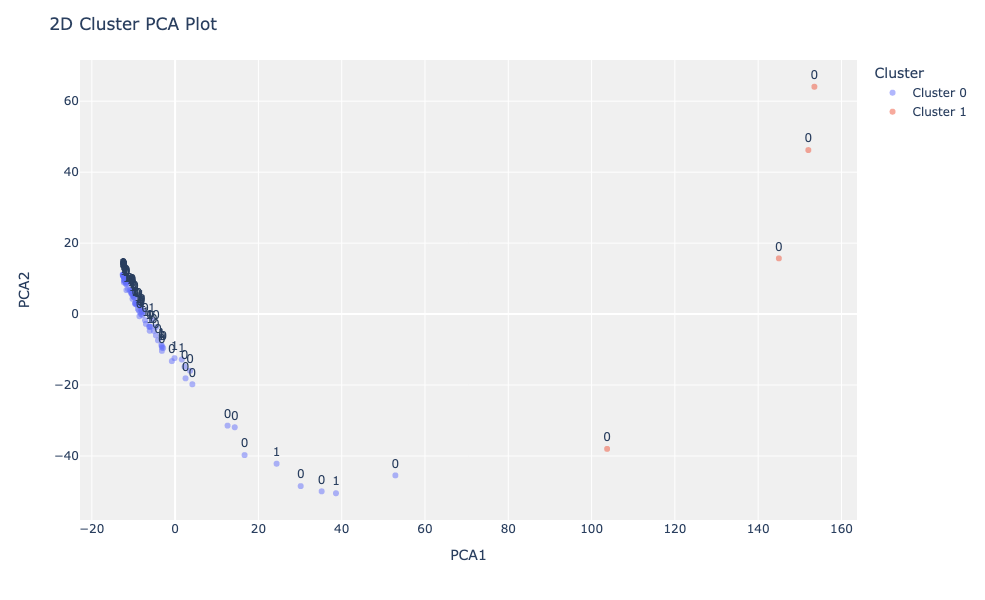}
            \label{fig:SRl}
    \end{subfigure}%
    \caption{PCA projection of K-means clustering}\label{fig:pca}
\end{figure}

In this section, we visualize the embeddings generated from the GCN layer using t-Distributed Stochastic Neighbor Embedding (t-SNE) \cite{ref_25} which is a feature reduction technique mitigating the\\
\vspace{-0.05cm}effects of the ``Curse of Dimensionality". We also conduct an unsupervised K-means clustering on the dataset and visualize the result using a dimensionality reduction algorithm Principal component analysis (PCA) \cite{ref_26}. Through comparison, we can find out the difference between applying unsupervised learning and semi-supervised learning. The results are showing in Fig \ref{fig:tsne_16} and \ref{fig:pca}.\par
\vspace{-0.6cm}
\vspace{-0.8cm}
We discover that there is a distinct boundary between label 1 and label 0 which are colored in blue and red in t-SNE projection of GCN embeddings in Fig \ref{fig:tsne_16}. However, there is no distinguishable separation of two K-means clusterings colored in red and blue from PCA projection Fig \ref{fig:pca}, in which labels are annotated in the graph as 1 and 0. It means that GCN semi-supervised method achieved much better performance compared to the unsupervised training by using only a small portion of labelled training data.

\section{Results and Discussion}\label{section:result}
In this section, we apply the model with the best performance to conduct predictions on the rest of 798 unlabelled occupation nodes to get the probabilities of their automated risk. Part of the results is shown in Table \ref{tab2}.
\vspace{-0.6cm}
\begin{table}
\centering
\caption{Top 10 Occupations in High Risk of Being Automated}\label{tab2}
\begin{tabular}{|l l|}
\hline
{\bfseries Occupation\hspace{8cm}}  & {\bfseries Risk\hspace{4cm}}\\
\hline
{File Clerks}  & {0.7002}\\

{Real Estate Brokers}  & {0.6968}\\

{Word Processors and Typists}  & {0.6943}\\

{Payroll and Timekeeping Clerks}  & {0.6941}\\

{Data Entry Keyers}  & {0.6940}\\

{Transportation Engineers}  & {0.6938}\\

{Credit Analysts}  & {0.6938}\\

{Insurance Appraisers, Auto Damage}  & {0.6938}\\

{Insurance Underwriters}  & {0.6938}\\

{Tax Examiners and Collectors, and Revenue Agents   }  & {0.6938}\\
\hline
\end{tabular}
\end{table}\par
\vspace{-0.5cm}
Our results show that the maximum probability of automated risk is 0.7002 which is the ``File Clerks''. We follow the definition in \cite{ref_1} and set 69\% as the cut-off point so that if the probability is greater than 69\% then this occupation has a risk of being automated. Then we obtain 230 occupations and a percentage of 25.3\% among all occupations are at risk of being automated. This result is lower than the experiments conducted by \cite{ref_1}\cite{ref_8}\cite{ref_9} which are about 50\% occupations are at risk and is also higher than \cite{ref_2}\cite{ref_5} which conducted a job-level estimation of 9\% automated risk. Considering the defects in occupation-level and job-level automated risk estimations, our result is quite reasonable to reach a median value between these two estimations since it predicts the occupation-level risk using granular task and skills data.\par
\begin{figure}
  \includegraphics[width=\linewidth]{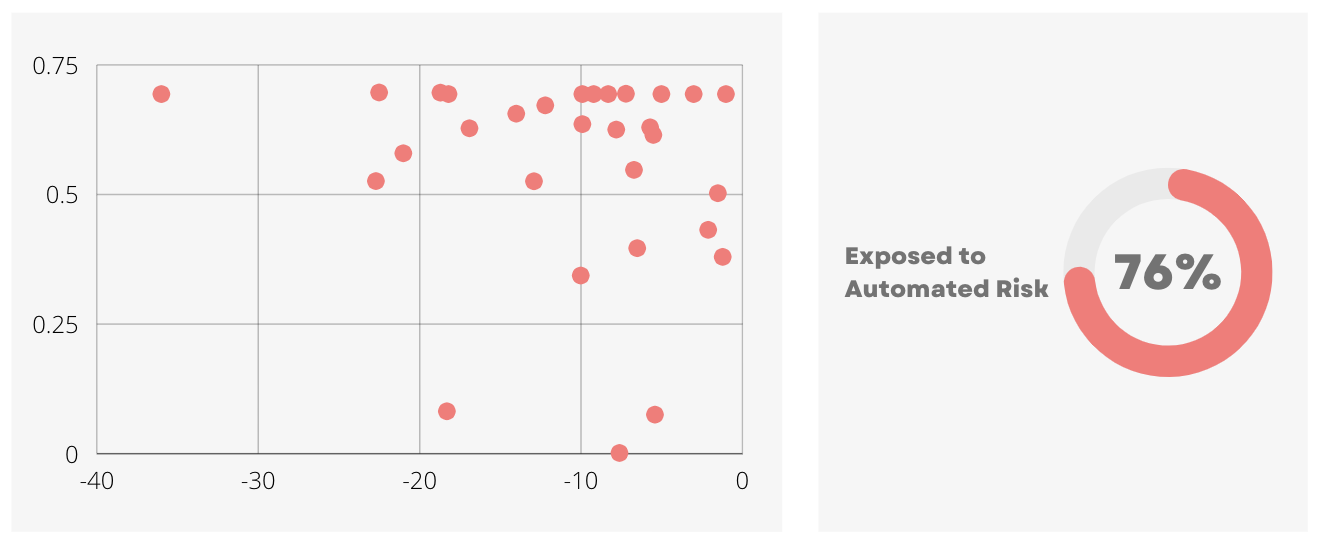}
  \caption{BLS Projected Declining Occupations in 2030}
  \label{fig:projected}
\end{figure}

 
To further validate our results, we compare them with ``U.S. Bureau of Labor Statistics (BLS) Occupations with the largest job declines, 2020 and projected 2030''. It projected out 29 occupations that have declining employments in 2030. The outcome is visualized in Figure \ref{fig:projected}.
\par
We assume that automated risk greater than 50\% would cause employment declining and we can find that 76\% of the BLS projected declining occupations have automated risks higher than 50\% except for ``customer service representatives", ``computer programmers", ``inspectors, testers, sorters, samplers and weighters" which have an automated risk probability of 0.1\%, 8.15\% and 7.5\% that are obviously automated safe but are projected declining in 2030. This discrepancy can be caused by rules used in expert assessment of the ground truth data. According to \cite{ref_1}, experts assess the risks based on three attributes of an occupation which are ``perception and manipulation", "creative intelligence tasks" and ``social intelligence tasks". i.e., ``customer service representatives" have tasks including persuading and dispute resolution which requires higher social intelligence and perception, therefore, they are not easy to be automated according to the analysis in \cite{ref_1}. BLS projection made predictions based on employment change in time series. And employment decline may be caused by various reasons e.g., supply and demand change \cite{ref_5}, other than automation technologies which mean the declining employment of ``customer service representatives" can be caused by the demand change, i.e. some the customers' basic demands can be satisfied by AI dialogue agents. That is to say, expert assessment based on occupations' attributes only can be biased since some occupations may still decline even if they have tasks and skills that automation technologies cannot perform because of the supply and demand change. Overall, the performance and evaluation demonstrate the effectiveness of our AOC-GCN method in occupation's automated risk prediction.

\section{Limitations and Future Work}\label{section:limit}
We acknowledge several limitations in our method and the results presented in this paper. Firstly, data limitations hindered the performance of our model. O*NET is the only globally admitted database in labour market, however, it only classified 1016 occupations, and this data size limited the performance of classification using machine-learning method. Moreover, each occupation is an aggregation of several jobs which still obfuscates the difference between jobs. Secondly, lack of ground truth data is also a barrier that inhibits predictions of automated risk. Currently, most of the ground truths are expert assessments based on occupation's attributes which are inevitably subjective to some extent. Expert assessments based on occupations' attributes can be biased without considering the supply and demand change. Moreover, there is no job-level or skill-level ground truth data which impedes predictions on a more granular level. Thirdly, predicting on static information cannot capture the technological change and labor trends since the employment trends and changing demand for specific tasks and skills might change faster than static O*NET data can capture. Therefore, our future work will focus on automated risk identifications on job-level which requires more granular ground truth data such as which skills or tasks decide a specific occupation being automated \cite{Dawson2021a}. Moreover, the study will be conducted on real-world datasets that capture technological change, employment trends and supply and demand change.

\section{Conclusion}\label{section:conclusion}
In this paper, we applied the AOC-GCN model to the Occupation-skill graph to identify the automated risk of a specific occupation. 
The initial node embeddings are generated from Word2vec and Doc2vec. 
GCN learns the graph structure and interactions between occupations. 
Extensive experiments and comparisons with real-world projections demonstrate the effectiveness of our method.
Our study paves the way for understanding where to concentrate upskilling efforts in the following decades \cite{Ahadi2022}. 
In the future, we will further extend our model to solve more granular-level tasks in labour market.

\end{document}